# Boundary-Induced Apparent Risk Aversion in Nonergodic Multiplicative Growth

Ling Zhang, [1,*] Boyan Xing,[2] Zhenyu She,[3] and Zixiang Xu[4]

[1]School of Economics, Guangdong College of Commerce, Guangzhou, China

[2]Department of Mathematical Sciences, Beijing Normal-Hong Kong Baptist University, Zhuhai, China

[3]School of Mathematics, The University of Edinburgh, Edinburgh, UK

[4]Department of Mathematics, College of William & Mary, Williamsburg, VA, USA

[*]Corresponding author: zhangling.thu@gmail.com

## Abstract

Finite multiplicative systems often cease to evolve when a lower continuation threshold is reached, whereas standard growth-optimal benchmarks assume uninterrupted continuation. We study a finite-horizon multiplicative process in which a fixed exposure is chosen ex ante and paths crossing an absorbing boundary are assigned a residual value. In the continuous-time limit the absorbed objective is available in closed form, and three results follow analytically. Costly absorption compresses exposure below the no-boundary Kelly fraction near the boundary; we prove that this holds strictly whenever the boundary is reachable within the horizon. The dependence on the horizon is non-monotone, and the optimum returns to a finite, cliff-independent long-horizon limit from below or from above according to an exact criterion, at which the threshold is the geometric mean of the residual value and the initial wealth, independently of drift and volatility. When interpreted through an unconstrained constant-relative-risk-aversion benchmark, this compression appears as elevated risk aversion. Exact lattice propagation confirms the analytical predictions without fitted parameters. Absorbing-boundary geometry can therefore generate state-dependent risk-averse-looking behavior without heterogeneous primitive preference parameters.

## 1. Introduction

Risk taking under uncertainty is commonly represented through expected utility. In the von Neumann–Morgenstern framework, preferences are represented by the expectation of a utility index, while the Arrow–Pratt measures relate local utility curvature to comparative risk aversion [1–3]. In portfolio theory, isoelastic or constant-relative-risk-aversion utility yields tractable and, under standard return assumptions, homothetic allocation rules across wealth levels and investment horizons [4–7]. This representational architecture is mathematically powerful and remains foundational.

For the present problem, however, it is useful to distinguish representation from mechanism. Utility curvature can rationalize a low exposure to a risky gamble, but it does not by itself identify which stochastic or institutional feature makes the exposure low in one state and high in another. Our aim is not to replace expected utility theory. It is to identify a complementary mechanism through which risk-averse behavior can arise even when the return process, objective function, and decision rule are held fixed.

A natural starting point is multiplicative growth. Kelly showed that repeated favorable bets admit an exposure that maximizes the exponential growth rate of wealth, and Breiman established strong long-run optimality properties for favorable repeated games [8,9]. Subsequent work extended the log-optimal criterion to more general investment environments [10,11]. Dynamics-based treatments have emphasized that in multiplicative processes the ensemble expectation of wealth need not coincide with the growth experienced along a single trajectory [12,13]. Related work has shown that a stabilising stochastic resetting mechanism can convert an otherwise non-ergodic multiplicative process into one with regime-dependent stationary behavior [14].

Nevertheless, the no-boundary growth benchmark omits a real-life feature of finite systems: continuation can depend on avoiding a lower threshold. First-passage theory characterizes the probability and timing of boundary crossing for stochastic processes, while classical gambling theory studies ruin and optimal play under specific payoff criteria [15–19]. Our question lies at their intersection. How does a finite absorbing threshold reshape the ex ante growth-optimal fixed exposure, and how would the resulting choice be interpreted by a model that omits the threshold?

By studying a finite-horizon multiplicative process with a lower absorbing boundary, we observe boundary-induced exposure compression. When absorption is costly, the same exposure has sharply different continuation consequences at different distances from the threshold. Near the boundary, an adverse finite-time sequence can terminate multiplicative growth; farther away, the same sequence may leave the path active. Optimal fixed exposure is therefore substantially below the no-boundary Kelly fraction near the threshold and approaches it as the boundary becomes dynamically irrelevant. To connect this result with conventional preference language, we map the boundary-constrained exposure into the CRRA coefficient that would rationalize the same choice in an unconstrained no-boundary benchmark. It rises sharply near a costly boundary. Thus, part of what a no-boundary model would encode as preference heterogeneity can instead be generated by heterogeneity in distance to a continuation threshold.

The central analytical contribution of this paper is that, in the continuous-time limit, this mechanism is exactly solvable. Writing the log-distance to the threshold as an arithmetic Brownian motion absorbed at the origin, the absorbed objective admits a closed form in terms of Gaussian integrals. Three consequences follow. First, compression is not a numerical observation but a theorem: whenever the boundary is reachable within the horizon, the optimal exposure lies strictly below the Kelly fraction, for every admissible residual value and every horizon. Second, the size of the boundary correction obeys a scaling law: it decays as a Gaussian in the single variable that combines initial distance, horizon and volatility, with an explicit inverse-horizon prefactor. Third, the horizon enters non-monotonically. Over a long horizon the liquidation cost is an $O(1)$ charge against an $O(T)$ accumulation of growth and is diluted away, so the optimum returns to a limit that does not depend on the residual value at all; whether it returns from below or from above is decided by an exact and remarkably simple criterion, and it involves neither the drift nor the volatility of the process.

Moreover, the boundary effect is not unidirectional. Its sign and magnitude depend on the value assigned after absorption. When $\rho$ is close to one, the residual state truncates additional downside while favorable paths retain multiplicative upside. A local near-boundary reversal can then occur, with optimal exposure exceeding the no-boundary Kelly benchmark. This is a structural limited-liability-like effect rather than evidence of a separate primitive taste for risk; related incentive effects near default and limited-liability boundaries are well known in corporate finance [20–23]. The continuum analysis sharpens this observation considerably. The contribution is therefore a boundary-based mechanism: finite absorbing thresholds convert the no-boundary Kelly fraction into a state-dependent optimal fixed exposure. Compression is generic and provable; reversal is a discreteness effect with a sharp criterion.

The remainder of the paper is organized as follows. Section 2 introduces the absorbing-boundary multiplicative model and the exposure-selection problem. Section 3 develops the continuous-time limit, derives the closed-form objective, and proves three results: the compression theorem, the horizon phase transition, and the reversal criterion. Section 4 presents the exact lattice method used to treat the discrete process. Section 5 reports the numerical results and their comparison with the analytical predictions. Section 6 discusses the relation to preference representation, nonergodic growth, first-passage dynamics, and limited-liability-like incentives. Section 7 concludes.

## 2. Model

### 2.1. Multiplicative dynamics and absorption

We consider a finite-horizon multiplicative growth process with a lower absorbing boundary. Time is discrete, $t = 0,1, \dots, T$. An agent begins with wealth $W_0$ and selects a fixed fractional exposure $f$ before the process starts. Wealth then evolves according to

$$W_{t+1} = W_t\,(1 + fR_t) \qquad (1)$$

where $R_t = +b$ with probability $p$ and $R_t = -a$ with probability $1 - p$. Here $a > 0$ and $b > 0$ denote the loss and gain coefficients, and we restrict the admissible exposure set so that $1 - fa > 0$. The same fixed exposure is used in every period; the object studied below is therefore an ex ante exposure-selection problem rather than a dynamic feedback policy.

The process is subject to an absorbing threshold $L > 0$. Once wealth reaches or falls below $L$, the trajectory is terminated and assigned a residual terminal value $S$, where $0 < S < L$. The first-passage time is $\tau_L = \inf\{t \geq 0 : W_t \leq L\}$, and terminal wealth under the absorption rule is $W_T^{\text{abs}} = W_T$ if $\tau_L > T$, and $S$ otherwise.

The distinction between $L$ and $S$ is structural. The threshold $L$ determines when continuation stops, whereas $S$ determines the terminal value assigned after absorption. A low $S$ creates a severe liquidation cliff; a value close to $L$ creates a shallow residual state. As will become clear, it is not $S$ itself but the cliff depth $c = \log(L/S)$ that governs the behavior of the optimum.

### 2.2. State variables and exposure-selection problem

The initial position of the trajectory is summarized by its log distance from the absorbing threshold, $d = \log(W_0/L)$, and liquidation severity by the residual ratio $\rho = S/L$, equivalently by $c = -\log \rho$. The horizon $T$ determines how many multiplicative steps the path must survive. The three quantities $d$, $c$ and $T$ therefore describe the initial buffer, the severity of absorption, and the duration of exposure to first-passage risk. For a given environment the agent selects $f$ to maximize expected absorbed log terminal wealth,

$$f^*(d, T, \rho) = \underset{f}{\operatorname{argmax}}\ \mathbb{E}\big[\log W_T^{\text{abs}}\big] \qquad (2)$$

The analysis does not introduce heterogeneity in this objective. Instead, it asks how a common objective and return process produce different optimal exposures when the initial distance, residual state, or horizon changes.

### 2.3. No-boundary growth benchmark

If the absorbing boundary is removed, the expected one-period log-growth rate is $g(f) = p\log(1 + fb) + (1 - p)\log(1 - fa)$, strictly concave on the admissible set (see Supplementary Material), with unique interior maximizer $f_K = [pb - (1 - p)a]/(ab)$ . In the symmetric benchmark $a = b = 1$ this reduces to $2p - 1$; with $p = 0.55$ the Kelly exposure is 0.10.

## 3. Continuous-time limit: closed form and three theorems

### 3.1. Diffusion formulation and closed-form objective

In the diffusion limit the process is geometric Brownian motion with fixed fractional exposure, $dW/W = f\mu\, dt + f\sigma\, dB$. Writing $X_t = \log(W_t/L)$, the state is arithmetic Brownian motion

$$dX = m\, dt + s\, dB, \qquad m(f) = f\mu - 1/2\, f^2\sigma^2, \qquad s(f) = f\sigma \qquad (3)$$

with $X_0 = d > 0$, an absorbing barrier at the origin, and terminal value $-c$ on absorption. The Kelly fraction is $f_K = \mu/\sigma^2$, the maximizer of $m$. By the method of images with drift the objective integrates in closed form. With the dimensionless groups $\alpha = d/(s\sqrt{T})$, $\theta = m\sqrt{T}/s$ and $\gamma = c/(s\sqrt{T})$,

$$\begin{aligned} J &= s\sqrt{T}\,\{(\alpha + \theta)\Phi(\alpha + \theta) + \varphi(\alpha + \theta) - e^{-2\alpha\theta}[(\theta - \alpha)\Phi(\theta - \alpha) + \varphi(\theta - \alpha)] \\ &\quad - \gamma\big[1 - \Phi(\alpha + \theta) + e^{-2\alpha\theta}\Phi(\theta - \alpha)\big]\} \end{aligned} \qquad (4)$$

with $\Phi$ and $\varphi$ the standard normal distribution and density. No simulation and no lattice enters (4): the dependence on the five raw parameters is carried by three groups. Two auxiliary quantities recur below. Writing $\kappa = 2md/s^2$, the eventual-ruin probability is $q(f) = \mathbb{P}(\tau < \infty) = e^{-\kappa}$, and we set $P(f) = \mathbb{P}(\tau \le T)$ and $R(f) = \mathbb{E}[(T - \tau)^+]$.

### 3.2. Compression: the optimum lies strictly below Kelly

Let $\Delta(f) = J(f) - J_0(f)$, where $J_0(f) = d + m(f)T$ is the unconstrained objective.

**Lemma 1**. $\Delta(f) = -c\, P(f) - m(f)\, R(f)$.

**Lemma 2**. For every $f > 0$, both $P$ and $R$ are strictly increasing in $f$.

**Theorem 1 (compression)**. Assume $\mu > 0$, $0 < S < L$, and that the boundary is reachable within the horizon. Then $f^* < f_K$ strictly, for every admissible residual value and horizon.

If instead the boundary cannot be reached, $\Delta \equiv 0$ and $f^* = f_K$ exactly; this degenerate case is the continuum counterpart of the inactive-boundary condition $d > -T\log(1 - fa)$ of the discrete model.

### 3.3. Horizon dependence: an exact phase boundary

Lemma 1 also delivers the large-horizon behavior. Since $R = \int_0^T \mathbb{P}(\tau \leq u)\, du$ and $\mathbb{P}(\tau \leq u) \to q$, one has $R = qT - \mathbb{E}[\tau; \tau < \infty] + o(1)$, whence

$$J(f, T) = T\, H(f) + G(f) + o(1), \qquad H = m(1 - q), \qquad G = d - cq + m\, \mathbb{E}[\tau; \tau < \infty] \tag{5}$$

The leading term identifies the infinite-horizon optimum $f_\infty(d) = \operatorname{argmax} H$, which is independent of $c$: over a long horizon the cliff cost is an $O(1)$ charge against an $O(T)$ accumulation of growth, and is diluted away. The subleading term $G$ decides from which side the optimum returns, and it simplifies remarkably.

**Lemma 3**. $\mathbb{E}[\tau; \tau < \infty] = (d/m)\, q$. Consequently $G(f) = d + (d - c)\, q(f)$.

**Theorem 2 (horizon phase transition)**. Let $A(d, c) = \lim_{T\to\infty} T[f^*(d; T) - f_\infty(d)]$. Then $\operatorname{sign} A = \operatorname{sign}(d - c)$. The optimum returns to its infinite-horizon limit from below when $c > d$ and from above when $c < d$. The boundary is $c = d$, equivalently $L^2 = SW_0$: the threshold is the geometric mean of the residual value and the initial wealth. It does not depend on $\mu$ or $\sigma$.

Two remarks are needed to state the result carefully. First, Theorem 2 governs the direction of the asymptotic approach, not the presence of a dip at any particular finite horizon. Numerically, an undershoot of $f_\infty$ persists somewhat into the region $c < d$, vanishing only near $c/d \approx 0.75$ at $d = 0.8$; that second boundary has no simple closed form and is reported numerically in Fig. 7. Second, on the critical line itself the leading coefficient vanishes and the approach is governed by higher-order terms. Both features are visible in Fig. 7(b).

### 3.4. When can the optimum exceed Kelly?

Path continuity enters Lemma 1 only through $X_\tau = 0$. Without it the same computation gives $\Delta = \mathbb{E}\big[\mathbf{1}_{\{\tau \leq T\}}\big(-c - X_\tau - m(T - \tau)\big)\big]$, whose integrand can be positive.

**Theorem 3 (reversal criterion).** An above-Kelly optimum is possible only if $\mathbb{P}(X_\tau < -c) > 0$, i.e. only if the process can overshoot the boundary by more than the cliff depth. For continuous paths $X_\tau \equiv 0$, so no reversal occurs for any $\rho < 1$.

The criterion also says where to look. Small $d$ makes large exposures optimal and hence overshoots deep, while large $\rho$ makes $c$ small and hence easy to exceed; both hold only in a narrow near-boundary, shallow-cliff corner, which is exactly where the lattice locates the effect in Section 5.4.

## 4. Exact lattice propagation for the discrete process

The continuum results of Section 3 describe the diffusion limit. To treat the discrete process itself — and, as it turns out, to expose an effect that the continuum cannot represent — we evaluate the absorbed objective by exact finite-horizon probability propagation rather than Monte Carlo simulation. For a fixed exposure $f$, define the gross multipliers $A = 1 + fb$ and $B = 1 - fa$. A lattice node reached after $t$ periods and $k$ favorable moves has wealth $W_0 A^k B^{t-k}$.

Let $P_t^A(k)$ denote the active probability mass at node $(t, k)$, after all mass that crossed the boundary at or before $t$ has been removed, and let $P^{\text{abs}}$ denote cumulative absorbed mass. From an active node the favorable and adverse branches carry mass $pP_t^A(k)$ and $(1 - p)P_t^A(k)$; any descendant with wealth at or below $L$ is transferred to $P^{\text{abs}}$, the rest to the next layer. After $T$ periods,

$$\mathbb{E}\left[\log W_T^{\text{abs}}\right] = P^{\text{abs}} \log S + \sum_k P_T^A(k) \log\left(W_0 A^k B^{T-k}\right) \qquad (6)$$

Because the binary tree recombines, each layer contains at most $t + 1$ nodes and the computation scales polynomially with $T$ rather than with the number of raw return sequences. The benchmark parameterization is $L = 100$, $S = 10$ ($\rho = 0.1$), $T = 50$, $p = 0.55$, $a = b = 1$, with 601 exposures on $[0,0.3]$ and 80 distances on $[0.05,3.0]$. Ties within $10^{-12}$ are broken toward the smaller exposure; this is a reporting convention, not a behavioral preference. Local refinements use 1201 exposures. Full conventions and the corresponding diagnostics are collected in the Supplementary Material.

For the symmetric benchmark the boundary-constrained exposure is mapped into a no-boundary CRRA coefficient through $\gamma_{\text{app}} = \log[p/(1 - p)] / \log[(1 + f^*)/(1 - f^*)]$. This is a benchmark-implied diagnostic, not a measurement of psychological preference. When $f^* = 0$ the raw value is infinite; any finite display cap is used only for plotting.

## 5. Results

### 5.1. Model geometry and survival probabilities

We begin by illustrating how the absorbing boundary changes the geometry of the process. Without a boundary, the benchmark gamble has a well-defined log-growth optimum. With a finite threshold, the same sequence of returns can have different consequences depending on the initial position of the path.

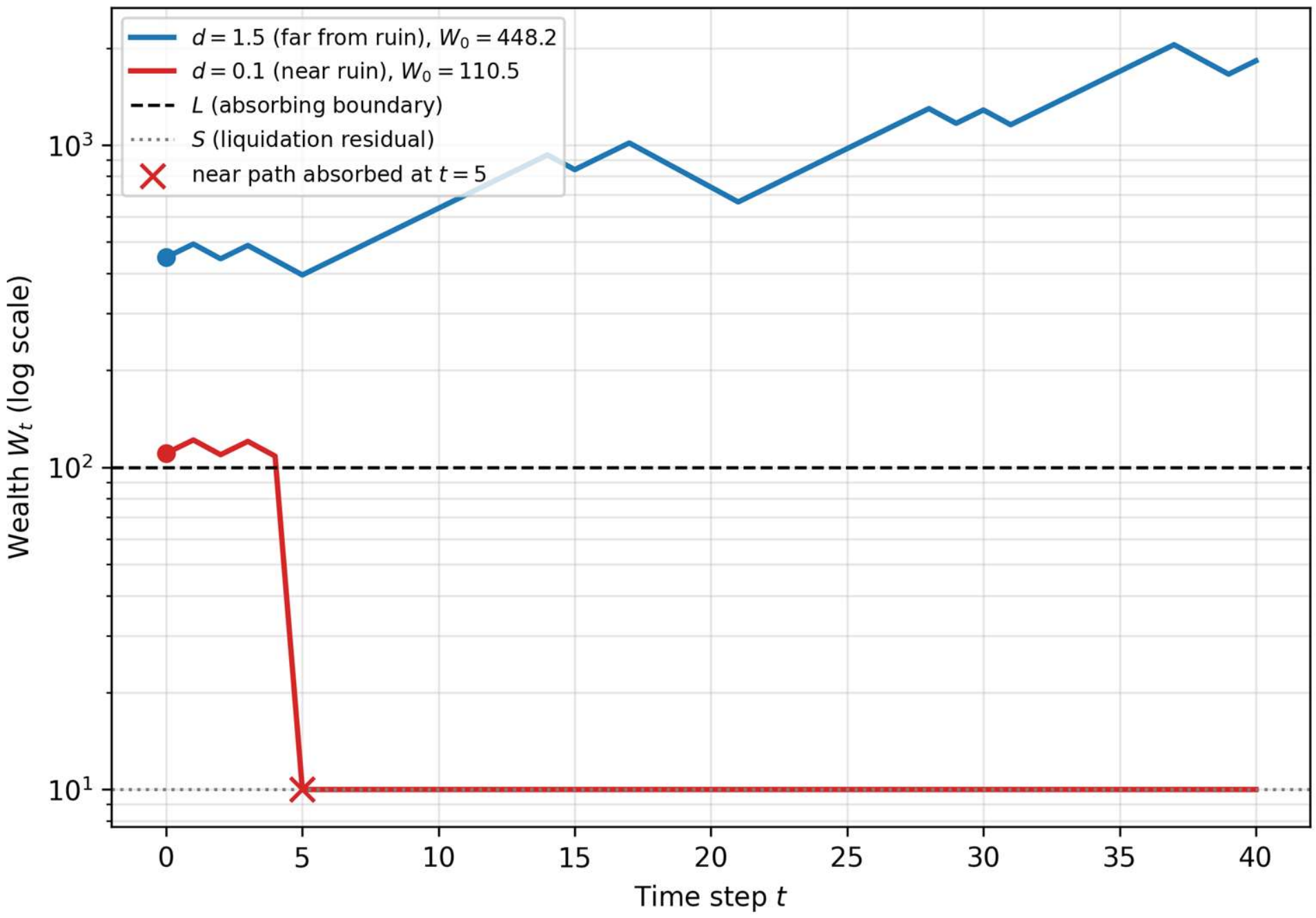

Fig. 1. Model geometry under a common return sequence and fixed exposure $f = 0.10$.

Figure 1 compares two trajectories exposed to the same realized shock sequence and the same fixed exposure. The path beginning farther from the threshold remains active, whereas the near-boundary path crosses $L$ and is assigned the residual value $S$. Figure 2 reports the end-horizon survival probability as a function of exposure for selected initial distances. Far from the boundary, survival remains close to one over a wide range of exposures; near the boundary it declines sharply as exposure increases. The same fixed exposure therefore produces different continuation consequences depending on the initial buffer, and this is the tradeoff that generates the state-dependent optimum.

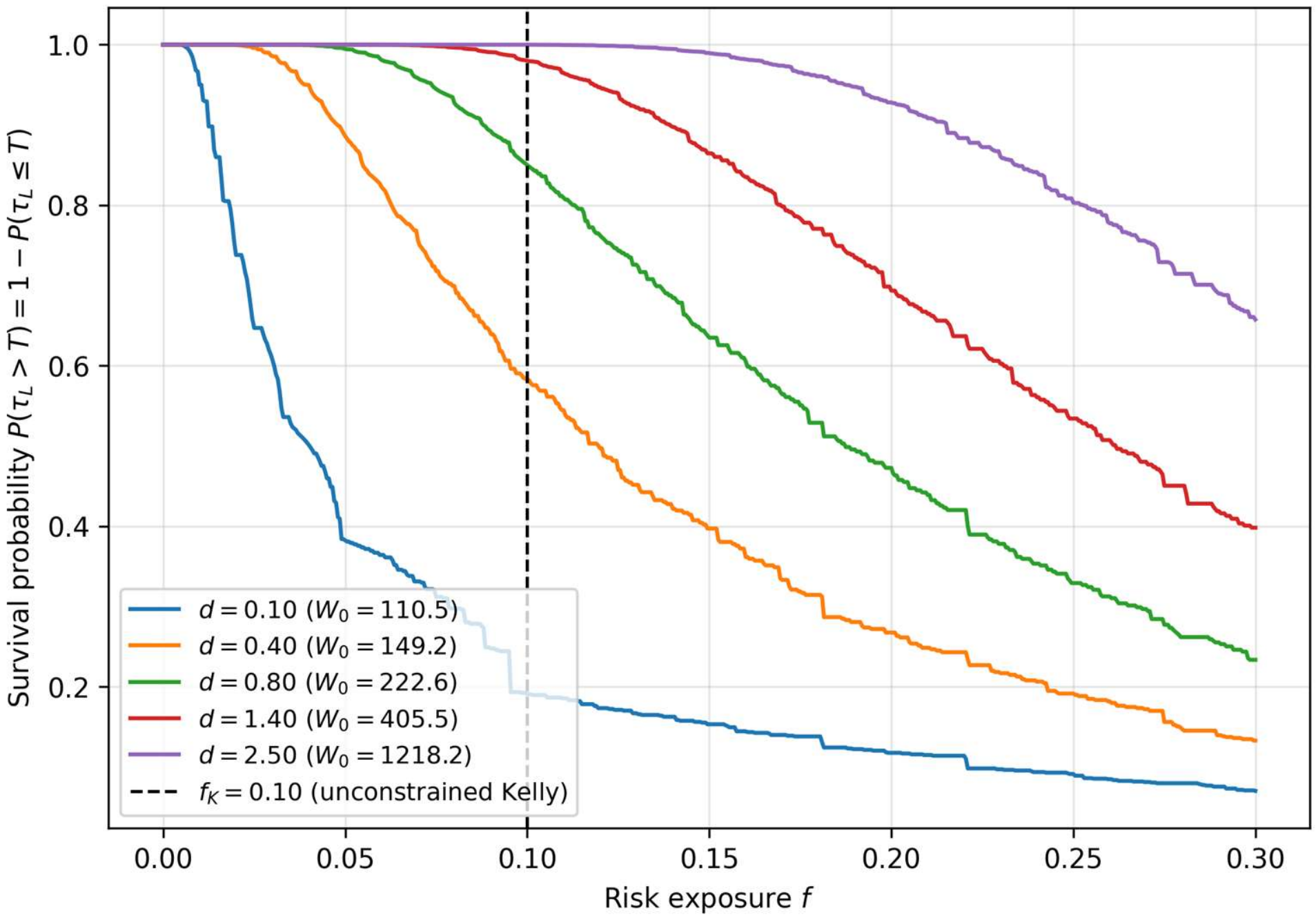

Fig. 2. End-horizon survival probability as a function of fixed exposure $f$ for selected initial log distances $d$.

## 5.2. Boundary-induced compression, and its analytical prediction

Figure 3 shows $f^*(d)$ under the benchmark parameterization. The dominant pattern is a strong boundary-induced compression: near the threshold the optimizer selects exposures far below the Kelly benchmark, and as $d$ increases $f^*$ rises and approaches $f_K$. The solid curve is the closed-form prediction (4) of Section 3, evaluated with the diffusion parameters implied by the binary benchmark; no parameter is fitted. The two independent computations agree to within the lattice grid resolution over the whole range, which simultaneously validates the lattice implementation and confirms that the discrete result is the diffusion result up to discretization corrections.

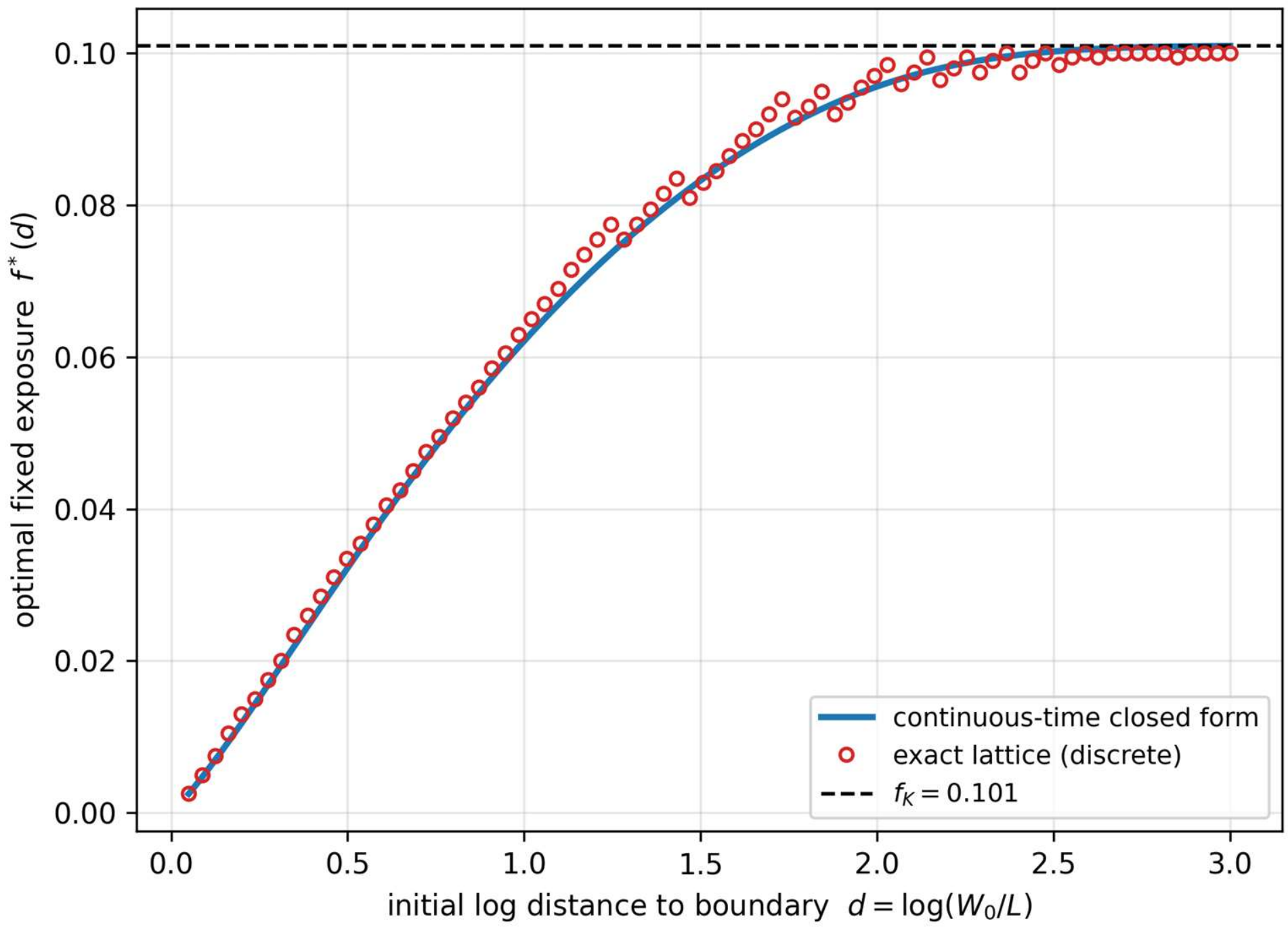

Fig. 3. Boundary-constrained optimal fixed exposure $f^*(d)$. Markers: exact lattice. Solid curve: continuous-time closed form, Eq. (4), with no fitted parameters.

The mechanism is path-geometric rather than preference-based. Increasing $f$ raises growth on favorable surviving paths but also raises the probability that adverse paths are transferred to the residual state; Lemma 1 makes this tradeoff explicit, since the boundary correction is exactly $-cP - mR$. All calculations use the same return process, objective and liquidation rule; only the initial distance changes. Lower exposure near the boundary therefore does not require heterogeneous primitive preferences.

Small local irregularities in the lattice curve should not be read as behavioral regimes: in a finite lattice, small changes in $d$ alter the discrete set of nodes that cross the threshold. That these irregularities are absent from the smooth closed-form curve confirms their discretization origin.

### 5.3. Apparent risk aversion under a no-boundary benchmark

The compressed exposure translates directly into the language of conventional risk aversion. Figure 4 shows that $\gamma_{\text{app}}$ is strongly state dependent: far from the boundary it approaches unity, the log-growth benchmark, while near the boundary the same underlying objective generates much smaller exposures, which a no-boundary CRRA model encodes as sharply higher apparent risk aversion. Because $\gamma_{\text{app}}$ is a monotone transform of $f^*$, the compression theorem implies $\gamma_{\text{app}} > 1$ wherever the boundary is reachable.

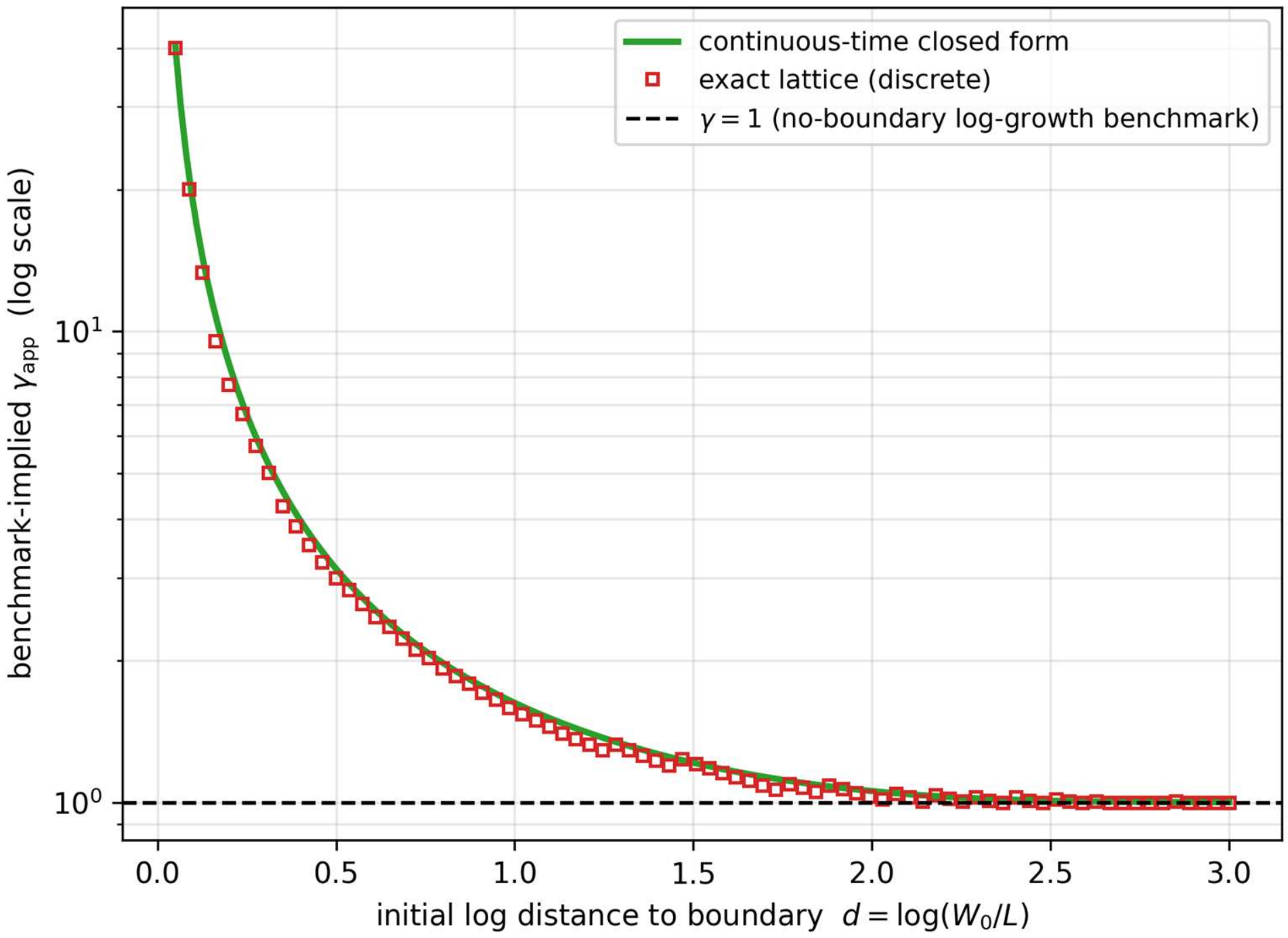

Fig. 4. Benchmark-implied apparent CRRA coefficient $\gamma_{\text{app}}(d)$, obtained by mapping the boundary-constrained exposure into a symmetric no-boundary CRRA benchmark. Markers: lattice; curve: closed form.

This mapping exposes an identification problem. A researcher who observes exposure but omits the continuation threshold would attribute low near-boundary exposure to high preference curvature. In the present model the high inferred $\gamma_{\text{app}}$ is a reduced-form representation of boundary proximity: variation in apparent risk aversion is generated by variation in state-space position. Concave utility can represent lower exposure, but it does not by itself identify why otherwise identical agents occupy different points on the exposure map.

### 5.4. Liquidation severity: compression, and a reversal that the continuum forbids

The boundary effect depends on the residual state as well as on distance. For low residual ratios, crossing is costly, the marginal cost of exposure is large, and $f^*$ sits well below $f_K$. As $\rho$ increases the cliff becomes shallower and the compression weakens. In the discrete model a qualitatively different pattern then appears: for $\rho$ close to one, a narrow near-boundary region exhibits $f^* > f_K$. High-resolution refinement (1201 exposures, 21 distances) confirms the effect for $\rho = 0.90$, 0.95 and 0.99, with the above-Kelly region widening as $\rho \to 1$.

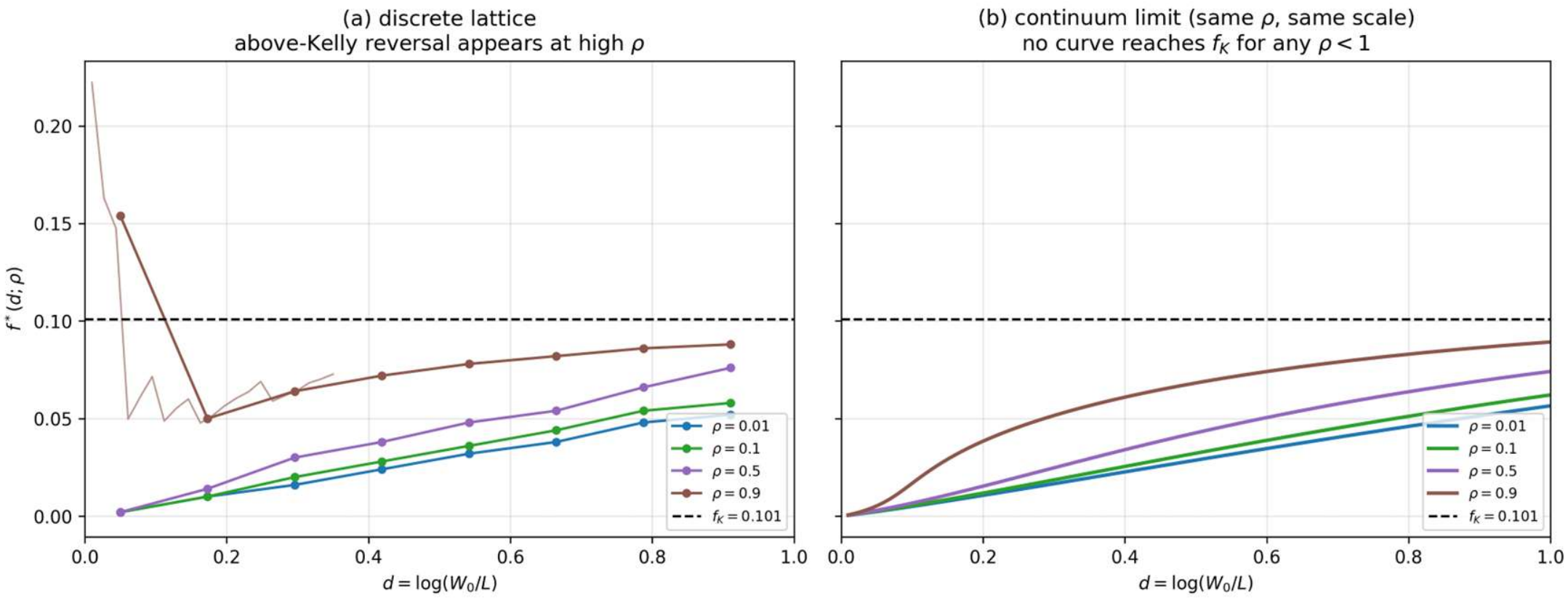

Fig. 5. Liquidation severity. (a) Exact lattice: at $\rho = 0.9$ the optimum rises above $f_K$ in a narrow near-boundary region. (b) The continuum limit at identical $\rho$, colors and scale: no curve reaches $f_K$ for any $\rho < 1$.

Figure 5 places the discrete and continuum results side by side, and the contrast is unambiguous: the reversal is absent from the continuum for every $\rho$ tested, up to 0.9999. Theorem 3 identifies the reason. Continuous paths meet the threshold exactly, so the value forgone on absorption is always the full cliff depth and absorption is always a strict loss. A discrete path, by contrast, can jump past the threshold: with $f = 0.222$ a single adverse step carries wealth from 105 to 81.9, so $X_\tau = -0.20$ while at $\rho = 0.9$ the cliff depth is only $c = 0.105$. Since $|X_\tau| > c$, absorption pays more than the path would otherwise have been worth, and the limited-liability-like incentive appears.

The reversal is therefore neither a numerical artifact nor a generic property of absorbing boundaries. It is a signature of discreteness, and the criterion of Theorem 3, $\mathbb{P}(X_\tau < -c) > 0$, predicts precisely where to look for it: at small distances, where large exposures are optimal and overshoots are deep, and at shallow cliffs, where the threshold to be exceeded is low. The same criterion implies that the effect should also arise in continuous-time models with jumps, or under discrete monitoring of an otherwise continuous process — settings in which a barrier can be crossed rather than touched.

### 5.5. Time horizon: a non-monotone dependence

In a multiplicative first-passage process the horizon is not a neutral scaling parameter: a longer path has more opportunities to reach the threshold. For fixed $f$ and $d$ the absorption events are nested, so $\mathbb{P}(\tau \leq T)$ is nondecreasing in $T$. This event inclusion is often taken to suggest that longer horizons uniformly deepen the compression.

This nesting is often taken to imply that longer horizons uniformly deepen the compression, and lattice calculations restricted to $T \lesssim 200$ appear to confirm it. The closed form shows that this reading is an artifact of the sampled range: the dependence on $T$ is genuinely U-shaped, and that range terminates near the minimum. The non-monotonicity is a property of the model, not of the lattice.

Figure 6 shows the full picture. Panel (a) plots $f^*(d;T)$ over an extended horizon range: the curves are ordered by $T$ up to a few hundred periods, and then the ordering reverses. Panel (b) plots the same quantity against $T$ directly, and the U-shape is explicit at every distance shown, with the minimum shifting to larger $T$ as $d$ increases. Panel (c) cuts across the phase boundary of Theorem 2 at fixed distance.

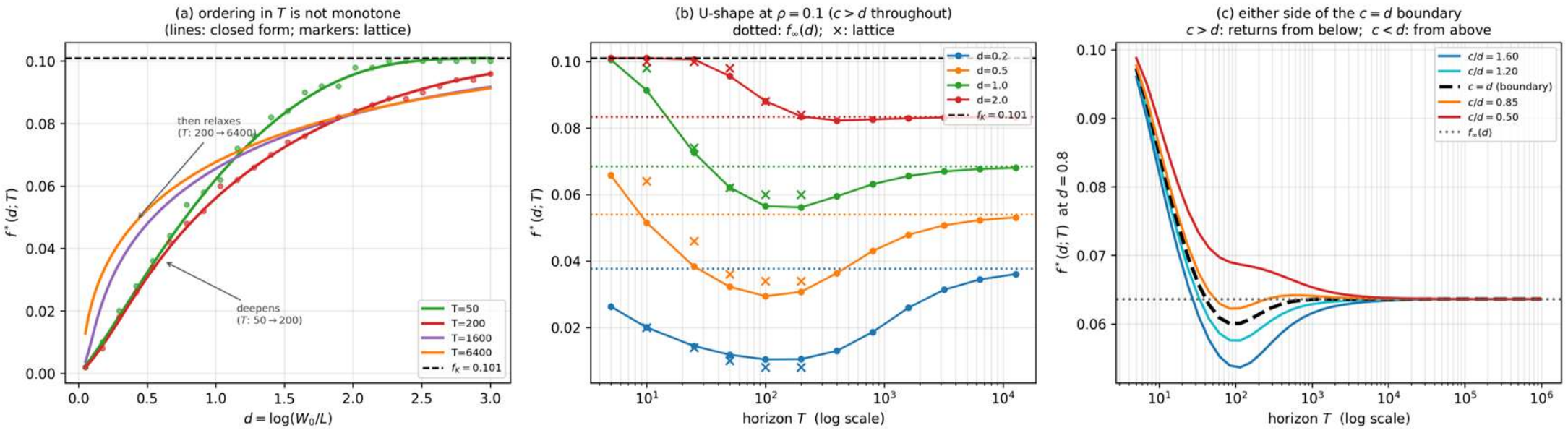

Fig. 6. Horizon dependence. (a) Optimal exposure against distance for several horizons; lines are the closed form, markers the lattice. The ordering in $T$ reverses beyond a few hundred periods. (b) The same optimum against $T$ on a logarithmic scale, showing a U-shape at every distance; dotted lines give the analytic infinite-horizon limit $f_\infty(d)$ of Eq. (7), crosses the lattice values, available only for $T \leq 200$. (c) Either side of the $c = d$ boundary at $d = 0.8$.

The continuum model explains the shape. For large $T$ the objective is dominated by $Tm(f)\mathbb{P}(\tau > T)$, and for arithmetic Brownian motion with positive drift the survival probability tends to the $T$-free limit $1 - \exp\big(-\kappa(f)\big)$ with $\kappa(f) = 2d\mu/(f\sigma^2) - d$. The maximization therefore converges to a horizon-independent problem,

$$f_\infty(d) = \underset{f}{\operatorname{argmax}}\, m(f)\big[1 - e^{-\kappa(f)}\big] \qquad (7)$$

so that $f^*$ first falls as first-passage risk accumulates, reaches a minimum, and then recovers toward $f_\infty(d)$. At $d = 2$ Eq. (7) gives $0.08332$ against $0.08330$ at $T = 25600$, and at $d = 1$ it gives $0.06841$ against $0.06822$. Crucially $f_\infty(d) < f_K$ for all finite $d$: compression survives the infinite-horizon limit, but it is approached non-monotonically. The lattice values, available only up to $T = 200$ because the recursion cost grows quadratically in the horizon, track the closed form over their whole range and sit on the descending branch.

Theorem 2 turns this observation into an exact statement. The direction from which the optimum returns to $f_\infty$ is decided by the sign of $d - c$ alone: for a deep cliff the optimum undershoots and recovers, while for a shallow one it descends monotonically. The critical line $c = d$ reads $L^2 = SW_0$ — the threshold sits at the geometric mean of the residual value and the initial wealth — and involves neither the drift nor the volatility.

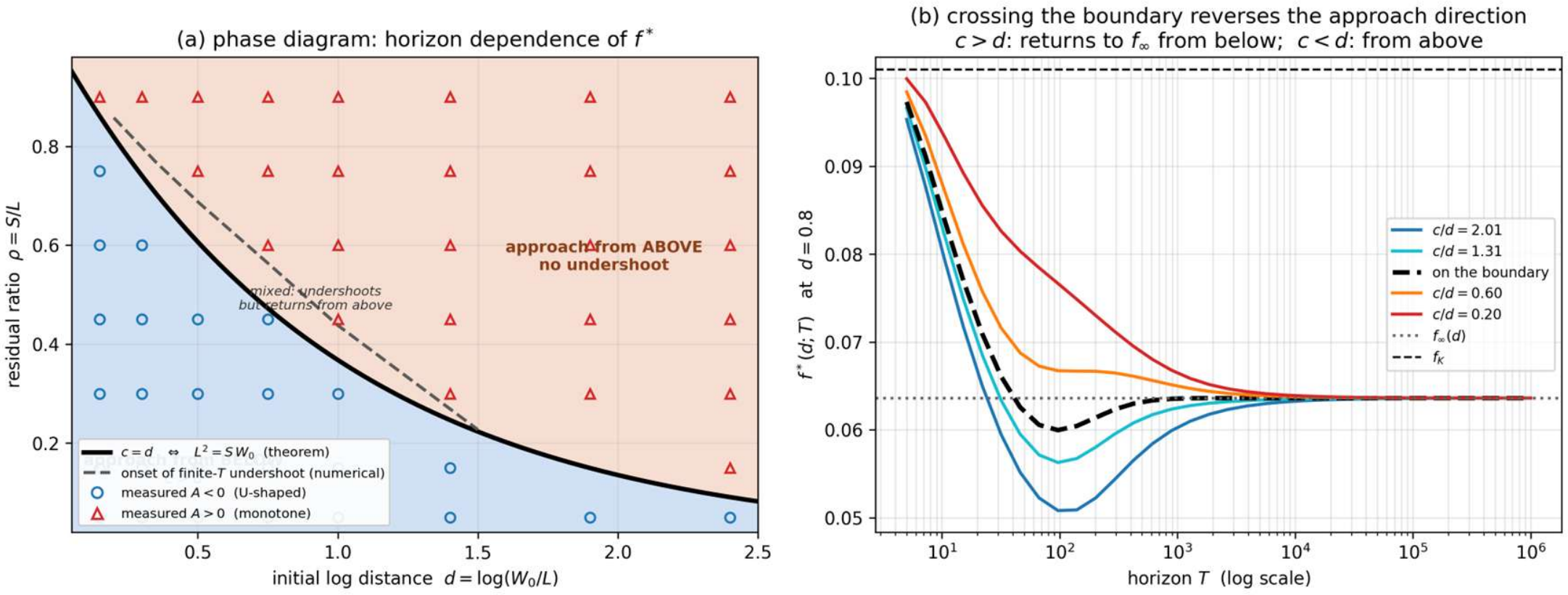

Fig. 7. Phase diagram of the horizon dependence. (a) The solid line is the exact boundary of Theorem 2, $\rho = e^{-d}$; markers show the numerically measured sign of $A$. The dashed line is the separate, numerically determined onset of a finite-horizon undershoot, which does not coincide with the theorem's line; between the two the optimum dips but still returns from above. (b) Cutting across the boundary at $d = 0.8$. On the critical line itself the leading coefficient vanishes and the residual shape is set by higher-order terms.

## 5.6. Scaling of the boundary correction

Finally, the closed form yields the rate at which the boundary ceases to matter. Expanding (4) for large distance shows that every correction term carries a Gaussian factor $\exp[-(\alpha+\theta)^2/2]$, and the first-order condition contributes an explicit inverse-horizon prefactor. Evaluating $\alpha + \theta$ at the Kelly exposure defines the scaling variable

$$z_K = \frac{d\sigma}{\mu\sqrt{T}} + \frac{\mu\sqrt{T}}{2\sigma} \qquad (8)$$

and predicts

$$f_K - f^* \simeq \frac{A}{T}\, z_K^2 \exp\left(-\frac{z_K^2}{2}\right) \qquad (9)$$

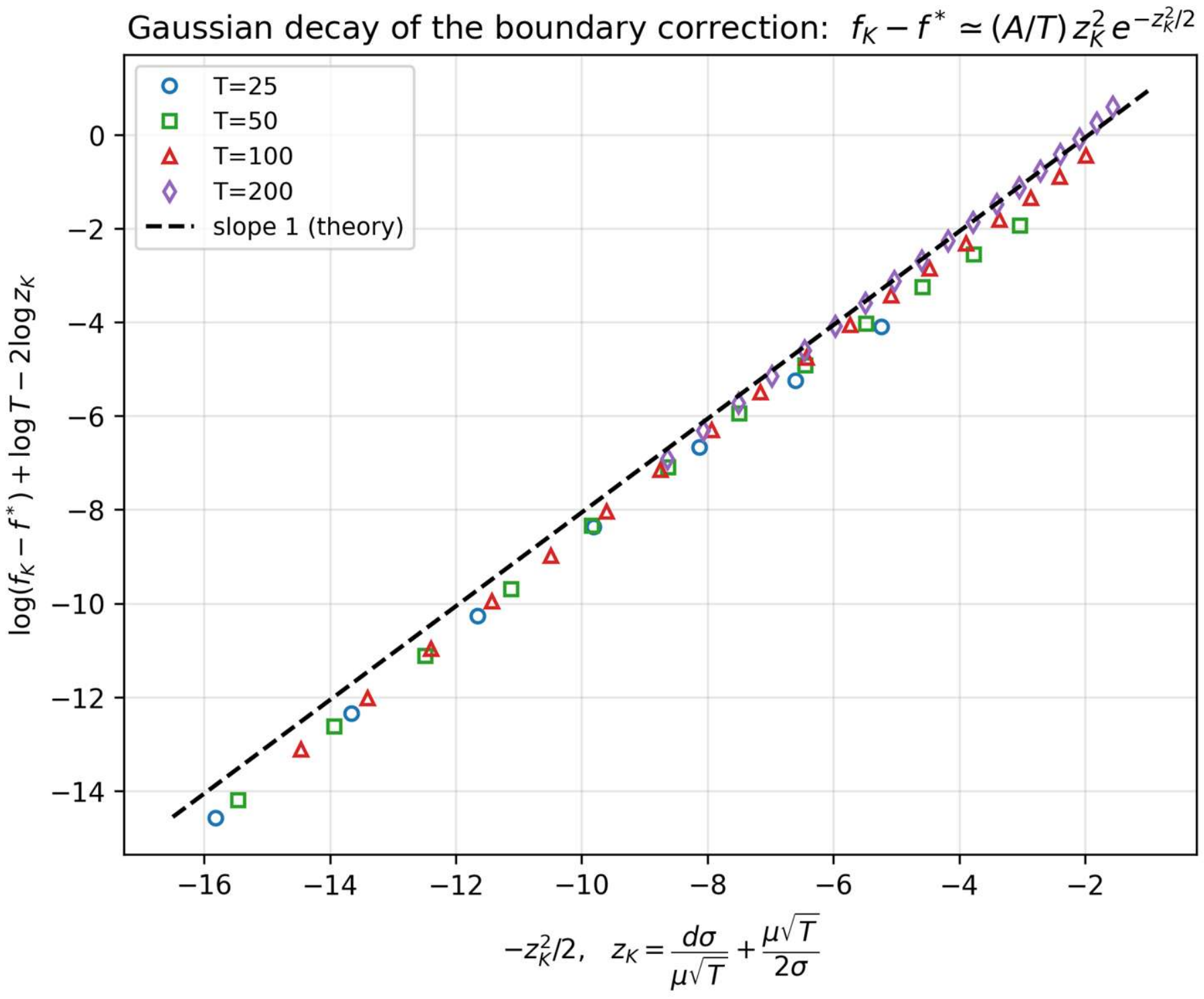

Fig. 8. Scaling collapse of the boundary correction. The 55 data points span four horizons and distances from 1.5 to 5, and fall on a common line of unit slope when plotted according to Eq. (9). The Gaussian coefficient fitted freely is 1.028, against the predicted 1.

Figure 8 tests (9) directly. The 55 points span an eightfold range of horizons and distances from 1.5 to 5 and collapse onto a single line of unit slope, with $R^2 = 0.996$; fitting the Gaussian exponent as a free parameter returns 1.028 against the predicted 1, and the polynomial prefactor exponent returns 1.879 against the predicted 2. The boundary correction is thus not merely small at large distance but Gaussian-small, which is why the recovery of the Kelly benchmark in Fig. 3 appears so abrupt. We note that (9) is an asymptotic statement supported by the expansion and by the collapse, not a rigorous error bound.

### 5.7. Summary of results

A finite absorbing boundary converts the no-boundary Kelly fraction into a state-dependent exposure map $f^*(d, T, \rho)$. Near a costly threshold, exposure is compressed because additional risk raises the probability of losing continuation value, and a no-boundary CRRA benchmark misreads this compression as elevated apparent risk aversion. The compression is provable, quantitatively predicted by a closed form, and governed at large distance by a Gaussian scaling law. Liquidation severity controls its strength; the horizon controls it non-monotonically, with an analytic infinite-horizon limit and an exact phase boundary. The one effect that does not survive the continuum — the above-Kelly reversal at shallow cliffs — is precisely the one that requires the process to jump.

## 6. Discussion

### 6.1. Representation and mechanism

The results distinguish between a representation of risk taking and a mechanism that generates it. In expected utility theory, lower exposure can be represented by greater curvature of the utility function. This representation is general and useful, but the curvature parameter is commonly treated as a primitive characteristic of the decision maker. By contrast, the present model holds the objective function and return environment fixed and changes only the agent's position relative to an absorbing boundary. The resulting variation in optimal exposure is generated by the stochastic environment rather than by an imposed variation in preference curvature.

This does not place the model outside expected utility theory; the control problem itself maximizes expected absorbed log terminal wealth. The point is one of interpretation. When two agents choose different exposures, the difference need not identify different primitive preferences if the agents face different continuation risks. The compression theorem sharpens this into a statement that holds for every admissible parameterization rather than for a particular numerical example: wherever the boundary is reachable, the observed exposure understates the agent's growth-optimal exposure, and the gap is a deterministic function of position, horizon and cliff depth.

## 6.2. Relation to growth optimality and first-passage dynamics

In the no-boundary process the Kelly fraction maximizes expected log growth. With a finite absorbing boundary the problem changes qualitatively: exposure affects not only the growth rate of active paths but also the probability that a path loses its continuation value. The Kelly fraction is recovered only when the boundary is dynamically irrelevant, and Section 3 makes the rate of that recovery explicit.

The present mechanism requires more than multiplicative nonergodicity alone. In an unconstrained multiplicative process, nonergodicity explains why time-average growth and ensemble-average wealth need not coincide. The absorbing boundary adds a second feature: some trajectories cease to participate in future growth. A related literature obtains regime-dependent behavior from stochastic resetting [14], where a trajectory is renewed rather than terminated; the structural contrast with absorption is that renewal preserves participation while absorption removes it, and it is removal that produces the compression studied here.

The relation to first-passage theory is complementary. Classical first-passage problems take the dynamics as given and characterize the probability or timing of boundary crossing. Here the first-passage structure enters an ex ante choice problem: the agent selects exposure before the path is realized, and that exposure changes both the distribution of active terminal wealth and the probability of absorption. Lemma 1 shows that the coupling between the two is exactly $-cP - mR$, a form that isolates the cliff cost from the forgone growth.

## 6.3. Liquidation residuals, overshoot, and the limits of the analogy

The residual ratio determines the economic meaning of the boundary. A low residual value makes crossing a liquidation cliff and produces strong compression. As the residual value rises, the marginal cost of exposure falls and the gap to the Kelly fraction narrows. In the continuum this narrowing is monotone but never crosses zero.

The reversal resembles limited-liability or risk-shifting incentives, but the analogy should not be pressed too literally. The model contains no debt contract, strategic transfer, or separate creditor; the effect follows solely from the terminal payoff geometry created by the residual state. What the criterion adds is a statement about when the analogy applies at all. The relevant comparison is not between $S$ and $L$, but between $S$ and the value the trajectory would have had at the moment of absorption. Any mechanism that separates these two — jumps, discrete monitoring, gap risk at a margin threshold — can generate the incentive, and any mechanism that does not, cannot.

### 6.4. Limitations and extensions

The model is intentionally minimal. Returns are independent and identically distributed and the agent selects one fixed exposure for the entire horizon. This makes the mechanism transparent and permits both the closed form and the exact lattice, but excludes features that matter in richer applications.

First, the fixed-exposure assumption differs from a dynamic feedback problem in which exposure responds to current wealth or remaining horizon. Furtherly, the boundary and residual value are deterministic and time invariant, whereas real liquidation and margin thresholds may depend on market conditions. Moreover, and most directly suggested by the present results, the reversal criterion points to jump-diffusion and discretely monitored barriers as the natural next setting: there the overshoot distribution is nondegenerate and $\mathbb{P}(X_\tau < -c)$ can be computed rather than merely bounded, which would turn the qualitative criterion into a quantitative prediction for the size and location of the above-Kelly region.

## 7. Conclusions

We have studied how a finite absorbing boundary changes growth-optimal exposure in a multiplicative stochastic process in which the agent selects one fixed exposure before a finite sequence of returns, and paths reaching a lower threshold are terminated and assigned a residual value.

In the continuous-time limit the problem is exactly solvable. The absorbed objective has a closed form in terms of Gaussian integrals, and the boundary correction reduces to $-cP - mR$, the cliff cost plus the forgone growth over the residual time. From this we proved that whenever the boundary is reachable within the horizon, the optimal exposure lies strictly below the Kelly fraction, for every admissible residual value and horizon; that the correction decays as a Gaussian in the scaling variable (8) with an inverse-horizon prefactor; and that the horizon dependence is non-monotone, with a closed-form infinite-horizon limit and an exact phase boundary at $L^2 = SW_0$.

Exact lattice propagation of the discrete process reproduces these predictions without fitted parameters, and adds one effect that the continuum forbids: a local above-Kelly reversal at shallow residual loss. We showed that this requires the process to overshoot the boundary by more than the cliff depth, a criterion that both explains the location of the effect and identifies it as a property of discrete or discontinuous dynamics rather than of absorbing boundaries as such.

Mapping the resulting choices into an unconstrained CRRA benchmark produces a sharply state-dependent apparent risk-aversion coefficient. These findings do not imply that preferences are irrelevant or that every instance of risk aversion is boundary induced. They establish a narrower identification result: in finite multiplicative systems, state-dependent exposure can arise from continuation geometry even when primitive preferences, beliefs, and return dynamics are held fixed.

## Data availability

The Reproducibility package files including the numerical data and custom code supporting the findings of this article are openly available according to request to corresponding author, complying with publication policies.